# Electrostatic Control over Temperature-Dependent Tunneling across a Single Molecule Junction

Alvar R. Garrigues[1], Lejia Wang[2], Enrique del Barco[1]* & Christian A. Nijhuis[2,3]*

[1] *Department of Physics, University of Central Florida, Orlando, Florida 32816 - USA*

[2] *Department of Chemistry, National University of Singapore, 3 Science Drive 3, Singapore 117543*

[3] *Centre for Advanced 2D Materials, National University of Singapore, 6 Science Drive 2, Singapore 117546*

(*) CORRESPONDING AUTHORS:

E.d.B: Tel.: (+1) 823 0755, e-mail: delbarco@ucf.edu

C.A.N: Tel.: (+65) 6516 2667, e-mail: christian.nijhuis@nus.edu.sg

**Understanding how the mechanism of charge transport through molecular tunnel junctions depends on temperature is crucial to control electronic function in molecular electronic devices. With just a few systems investigated as a function of bias and temperature so far, thermal effects in molecular tunnel junctions remain poorly understood. Here we report a detailed charge transport study of an individual redox-active ferrocene-based molecule over a wide range of temperatures and applied potentials. The results show the temperature dependence of the current to vary strongly as a function of the gate voltage. Specifically, the current across the molecule exponentially increases in the Coulomb blockade regime and decreases at the charge degeneracy points, while remaining temperature-independent at resonance. Our observations can be well accounted for by a formal single-level tunneling model where the temperature dependence relies on the thermal broadening of the Fermi distributions of the electrons in the leads.**

Tunnelling of electrons across molecular junctions depends on both temperature and applied bias. So far only a few systems have been reported that switch the mechanism of charge transport as a function of bias and temperature, resulting, for example, in good molecular diodes [1-4] or enabling long range charge transfer along molecular wires [5-9] or biomolecules [10-13]. Typically, such measurements display a transition between a temperature insensitive regime, governing conduction at sufficiently low temperatures, and a thermally-assisted transport regime, characterized by an exponential dependence with temperature. However, neither the origin of the thermal excitation nor the nature of the tunnelling process (coherent vs. incoherent) are fully understood in solid state junctions. For example, anomalous temperature-independent long-range (up to 16 nm) charge transport phenomena have been reported which cannot be straightforwardly explained with current theories [6,12,13]. Hence, it is important to deepen our understanding of how temperature influences charge transport phenomena in molecular junctions and how this information can be used to improve their electronic characteristics.

Usually, temperature-independent conduction in molecular junctions is associated with coherent tunnelling involving the molecular level nearest in energy to the Fermi-level of the leads (*i.e.*, the highest occupied molecular orbital (HOMO) or the lowest unoccupied molecular orbital (LUMO)) [5,14,16,17]. In this regime, the contact time of the charge carrier with the molecule is smaller than the intramolecular dephasing times. In contrast, an exponential increase of current with temperature is associated with an incoherent process, where the charge spends enough time in the molecule to allow decoherence [14]. In a wet electrochemical environment, where the current (*I*) increases with temperature (*T*) following the Arrhenius law ($I = I_0 e^{-E_a/k_B T}$, where $k_B$ is the Boltzmann constant and $I_0$ is a pre-exponential factor), the physical meaning of the activation energy $E_a$ is well-understood. It relates to thermal excitations of molecular vibrational modes and reorganization of neighbouring solvent molecules, (*i.e.*, inner- and outer-sphere reorganization processes, respectively [15]). In solid state molecular junctions, however, the absence of solvent molecules leaves the physical meaning of $E_a$ unclear [18]. Without intermolecular interactions, thermally activated interactions are restricted to coupling of the charge carrier to vibronic states of the molecular bridge, although the associated activation energies are often too low (<0.1 eV) to explain the observations. In this scenario, the value of $E_a$ depends also on charge image effects and the corresponding re-normalization of the molecular energy levels in the junction as a result of proximity with the metallic electrodes [19-21]. Thermal broadening of the Fermi occupation distribution of electrons in the leads and its overlapping with the molecular levels has been invoked as a cause of thermally activated conduction through a single molecule tunnel junction [22,23].

By far most temperature dependent charge transport studies have been conducted with two-terminal junctions based on self-assembled monolayers (SAMs) or single molecules contacted by conductive probes. These approaches so far only made it possible to investigate a limited range of temperatures close to room temperature and lack a gate electrode [1,24-30]. Although molecular single electron transistors (SETs) obtained via mechanical or electromigration techniques have a gate electrode enabling full control over the energy level alignment of the system, these junctions are normally kept at cryogenic temperatures to ensure stability [22,31-36]. It is therefore necessary to expand such temperature-dependent measurements in individual molecules by studying molecules that have been well characterized with other techniques (*e.g.*, SAM-based junctions) for comparative analysis.

Here we report electromigrated three-terminal junctions with single molecules where the conduction level is well localized within the molecule and isolated from the electrodes. We use a broad range of source-drain bias and gate voltages to study the temperature dependence of the junction across three conduction regimes: i) the Coulomb blockade regime (where the molecular level lies outside the conduction-bias window); ii) at the charge degeneracy point (where the level lies in between the electrostatic potential of the leads); and, iii) at resonance (where the molecular level matches the energy of one of the leads).

**Results**

**Junction fabrication**. Figure 1a shows schematically the single molecule transistor with a S-$(CH_2)_4$-Fc-$(CH_2)_4$-S (Fc=ferrocene) molecule used in this study (see Supplementary Methods). Similar molecules have been systematically studied in SAM-based junctions [1-3, 37-40], showing that at least three $CH_2$ units between the Fc and thiolate-metal bond are needed to localize the HOMO on the Fc unit [3]. In our experiments, the molecule is connected to the source and drain electrodes via a metal-thiolate bond and the redox-active Fc unit is well separated from the leads by two -$(CH_2)_4$- tethers to avoid delocalization of the HOMO, which is centred at the Fc. Figure 1b shows an SEM image of a three-terminal SET fabricated by means of electron-beam and optical lithography [41] (see Methods section for details).

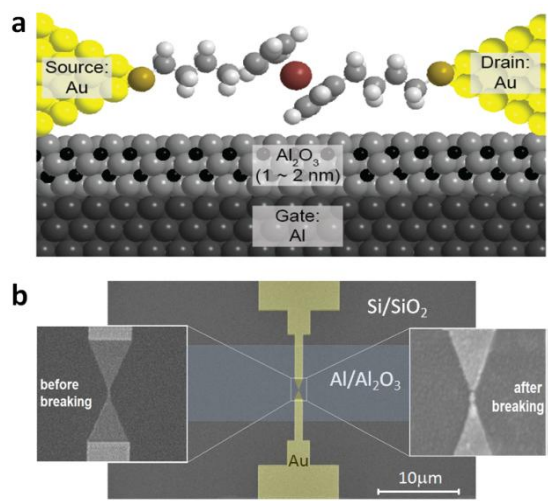

**Figure 1 | The single molecule junctions. a**, Schematic of a S-$(CH_2)_4$-Fc-$(CH_2)_4$-S molecule bridging the nanogap between the two bias leads of a SET. **b**, False color scanning electron microscopy (SEM) image of one of our SETs showing the gold nanowire on the Al/$Al_2O_3$ back gate. The scale bar is 10 μm. The insets show the thinnest part of nanowire before and after the feedback-controlled electromigration-induced breaking. The molecule completes the device by bridging the resulting nanogap during deposition from solution.

**Charge transport measurements**. Figure 2a shows the differential conductance ($dI/dV$) of the molecule at 80 K as a function of source-drain $V_{sd}$ (-150,+150 mV) and gate $V_g$ (-2.5,+3.5 V) voltages. Closed diamonds characteristic of single electron transport are clearly observed, with resonant transport excitations crossing at zero bias at two visible degeneracy points ($V_g$ = 0.25 and 2.5V), separating the Coulomb Blockade areas where no current flows through the molecule (dark blue areas). Figure 2b shows the source-drain current, *I*, as a function of gate voltage for different bias voltages corresponding to the horizontal color-coded dashed lines in Fig. 2a (the solid lines are fits to the model described below). The two consecutive charge degeneracy points start to intermix for the highest bias voltages ($V_{sd}$ = 50 and 70mV). The ratios of capacitive coupling *C* between Fc and the gate (g), source (s), and drain (d), electrodes can be extracted from the slopes of the resonant transport excitations (indicated by the tilted, dashed grey lines in Fig. 2a). The ratios are $C_g : C_s : C_d$ = 1 : 11 : 13, which determine a gate coupling parameter of $g_c \sim 0.05$, characteristic for a thin $Al_2O_3$ layer separating the molecule from the gate [22]. From these observations we conclude that the Fc unit is localized in the middle of the junction ($C_s \approx C_d$) and the HOMO energy level can be efficiently gated.

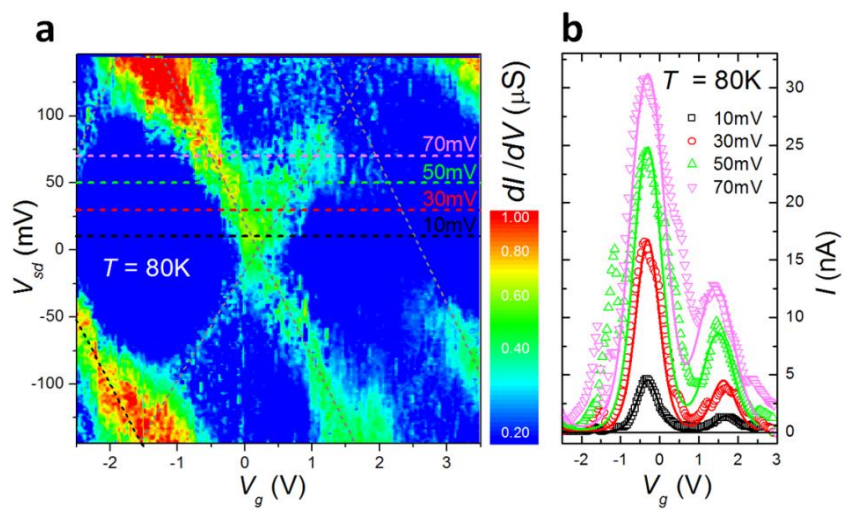

**Figure 2 | Charge transport at 80 K. a**, Differential conductance of a junction with S-$(CH_2)_4$-Fc-$(CH_2)_4$-S at $T$ = 80 K. The color code represents the conductance ($dI/dV$) through the SET as a function of $V_{sd}$ and $V_g$. The grey dashed lines indicate the main resonant excitations, crossing at two visible charge degeneracy points at $V_{sd}$ = 0 (*i.e.*, $V_g$ = 0.25 and 2.5V) and separating the Coulomb blockade areas. A third charge degeneracy point is estimated to lie around -3.5V. **b**, Current vs. gate voltage for four different bias voltages (10, 30, 50 and 70 mV). Symbols represent experimental values and solid lines are fits to the single-level tunneling transport model (Equation 1) using the parameters given in Table 1. The color code corresponds to the horizontal dashed lines in panel (**a**). The molecule most likely changed its conformation within the SET electrodes between both measurements resulting in a shift in the position of the degeneracy points (new positions: $V_g$ = -0.3 and 1.7V).

**Temperature dependent transport measurements**. Figure 3a shows a 3D plot of $I$ vs. $V_g$ and $T$ for $V_{sd}$ = 10 mV. The data were collected as the gate voltage was continuously swept from -2.5 to +2.5 V at each temperature, and the process repeated at different temperatures from 80 to 220 K. Similar results obtained for a second device are given in Supplementary Note 1 and Supplementary Figure 1 and 2, as well as a discussion of the reversibility of the temperature behavior of the molecular junction in the Supplementary Note 3 and Supplementary Figure 4. Fig. 3b is the corresponding theoretical plot using the single-level model described below. From these data, we can identify three transport regimes, each of which with a distinct dependence on the temperature as indicated by the arrows: *i*) the Coulomb blockade regime, with $I$ increasing exponentially with increasing $T$ (blue arrows); *ii*) the charge degeneracy points, with $I$ decreasing with $T$ (red arrows); and, *iii*) the resonant regime (when the molecular level matches the electrostatic potential of one of the leads), with *I* constant with *T* (black arrows).

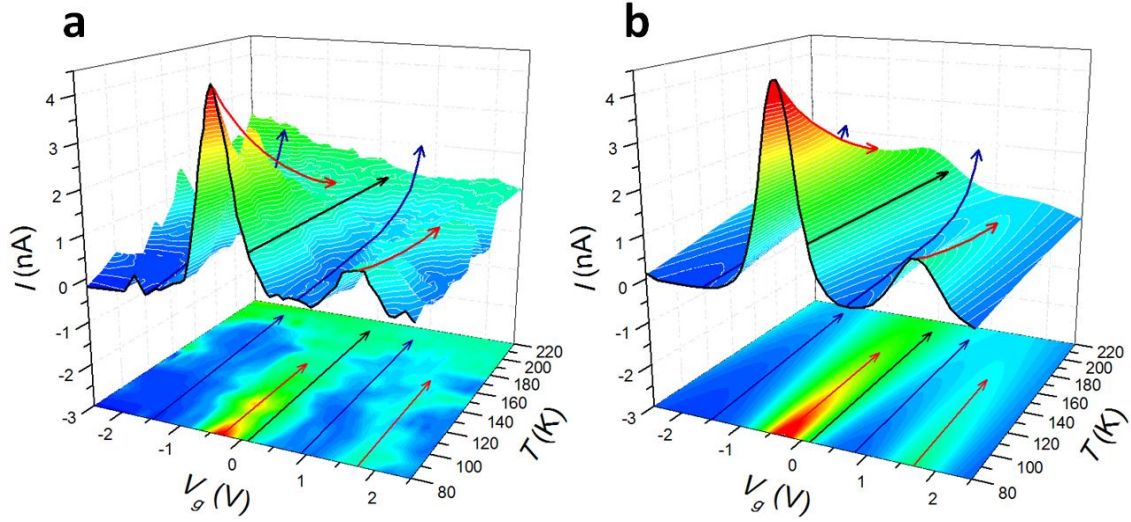

**Figure 3 | Variable temperature charge transport. a**, 3D plot of the evolution of the tunnel current through a S-(CH$_2$)$_4$-Fc-(CH$_2$)$_4$-S junction vs. gate voltage as the temperature is increased from 80 to 220 K for $V_{sd}$ = 10 mV. The evolution of the two charge degeneracy points (-0.3 V and 1.7 V), whose magnitude decreases with increasing temperature, can be appreciated following the red arrows in both the 3D data and its 2D horizontal projection in the $V_g$-*T* plane. Similarly, the increase with temperature in the Coulomb blockade areas can be seen by following the blue arrows. The black arrow shows the case for which the molecular level matches the electrostatic potential of one of the leads (resonance), a situation leading to almost negligible temperature dependence. **b**, Corresponding response of the junction as calculated from the single-level tunneling transport model in equation (1) using the parameters given in Table 1.

Measurements of the temperature dependence of the current were repeated at different bias values ($V_{sd}$ = 10, 30, 50, and 70 mV). Figure 4 shows the corresponding contour plots for each bias voltage and how the transport excitations at the degeneracy points broaden as a function of bias. The magnitude of *I* exhibits an overall increase as the value of $V_{sd}$ increases, as can be seen from the color-code level bars. Still, the distinct transport regimes and their respective temperature dependencies remain present for all examined $V_{sd}$ values.

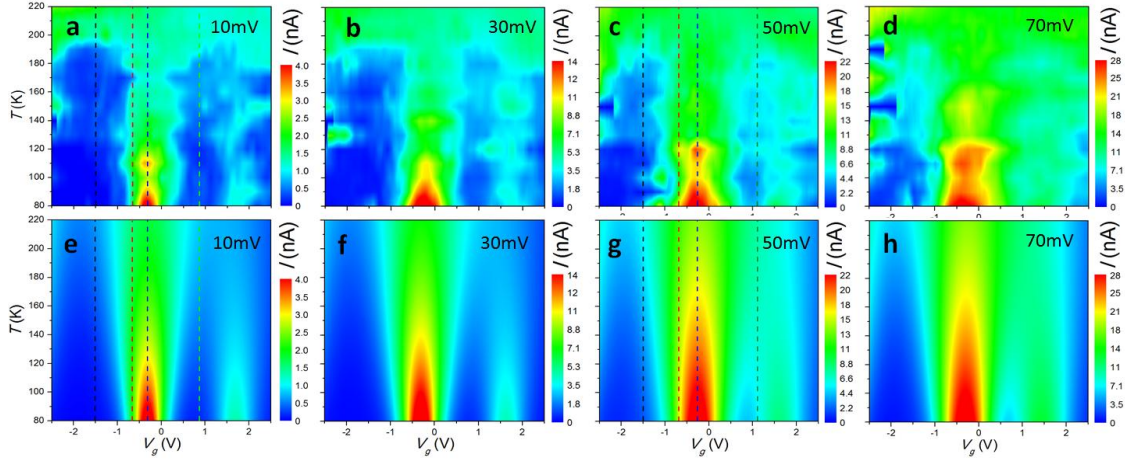

**Figure 4 | Variable temperature and source-drain voltage charge transport.** Contour color-code plots of the current as a function of the gate voltage and temperature for both experimental data (**a-d**) and calculations (**e-h**) using the single-level tunneling transport model in equation 1 for $V_{sd}$ = 10, 30, 50 and 70 mV. The vertical dashed lines in panels **a**, **c**, **e** and **d** indicate data and calculations presented in Figure 5.

**Single-level transport model**. We have used a single-level tunneling model to fit our experimental data using parameters that can be experimentally verified based on the Landauer formalism [42,43]. The conduction through the junction can be described as sequential tunnelling (either coherent or incoherent) from the right electrode into a single molecular level at a rate $\gamma_R$, and from the molecule into the left electrode at a rate $\gamma_L$, one electron at a time. The molecular level broadens to a width $\gamma = \gamma_L + \gamma_R$ as a result of its interaction with the respective electrodes. Sequential coherent tunnelling through a single-level molecular junction results in electrical current given by the expression derived by Jauho, Wingreen, and Meir in 1994 [44] using a fully coherent formulation based on the Keldysh Green's function formalism:

$$I = \frac{q}{h} \int_{-\infty}^{\infty} dE\, D_\varepsilon(E) \frac{\gamma_L \gamma_R}{\gamma_L + \gamma_R} [f_L(E) - f_R(E)] , \qquad (1)$$

where $q$ is the charge of the electron and $h$ is the Planck's constant. The molecular level is described by a broadened density of states $D_\varepsilon$ in the shape of a Lorentzian centred at the molecular level energy $\varepsilon$, as follows:

$$D_\varepsilon(E) = \frac{\gamma/2\pi}{(E-\varepsilon)^2 + (\gamma/2)^2}, \tag{2}$$

with the position of the molecular level with respect to the electrostatic potentials, $\mu_{L,R}$, of the transistor leads given by $\varepsilon = \mu_R(V) + \varepsilon_0 + \eta q V_{sd}$. Here, $\eta = V_R/(V_L + V_R)$ is the dimensionless division parameter giving the ratio of the voltage drop between the molecule and the right electrode with respect to the total voltage drop in the junction. In other words, $\eta$ represents the degree of symmetry of the electrostatic potential profile across the junction, with $\eta = 0.5$ representing a completely symmetric molecule and $\eta \sim 0$ or 1 accounting for a highly asymmetric potential drop at both sides of the molecule. In our junctions, $\eta = 0.5$ as determined from the similar slopes of the transport excitations in Fig. 2a (*i.e.*, $C_s \approx C_d$). Finally, the effect of temperature is included in equation (1) through the Fermi distributions of the electron occupation in the leads, given by:

$$f_{L,R}(E) = \frac{1}{1 + \exp\left[(E - \mu_{L,R})/k_B T\right]}. \tag{3}$$

It is important to realize that for high enough temperatures (*i.e.*, $\gamma \ll k_B T$), equation (1) reduces to $I = 2q\,\gamma_L \gamma_R/(\hbar[\gamma_L + \gamma_R])[f_L(E) - f_R(E)]$, which is the exact result one would obtain using rate equations to describe incoherent tunneling through a single-level in a molecular junction [42,43]. Indeed, in the case that only one molecular level is involved in the conduction through the junction (which is the case in the absence of intramolecular charge relaxation), coherent and incoherent descriptions of the process become indistinguishable for all temperatures if one also considers level broadening when solving the rate equations. Note that vibrational excitations of the molecule itself would open new channels for conduction, however, the low energies associated with these excitations (<30 meV) [45] allow renormalizing the molecular level by simply increasing $\gamma_{L,R}$ in equation (2).

**Interpretation of the temperature dependencies**. To explain the temperature behavior of the current in the junction, we fitted all data to the model given by equation (1) using the parameters listed in Table 1 (solid lines in Figure 2b, and Figures 3b and 4e-h). Three molecular levels were employed to account for the three degeneracy points observed within the window of electrical potentials. In essence, only two free-fitting parameters are used per molecular level to fit the results with Equation (1), *i.e.*, the respective tunneling rates $\gamma_{L,R}$. The energy levels, $\varepsilon$, are independently determined from the experimental data in Figure 2a.

**Table 1** | The fitting parameters to model the experimental data.[a]

| $V_{sd}$ (mV) | $\varepsilon_1$[b] (mV) | $\gamma_{1L}$ (mV) | $\gamma_{1R}$ (mV) | $\varepsilon_2$[b] (mV) | $\gamma_{2L}$ (mV) | $\gamma_{2R}$ (mV) | $\varepsilon_3$[b] (mV) | $\gamma_{3L,R}$ (mV) |
|---|---|---|---|---|---|---|---|---|
| 10 | 25 | 1 | 0.120 | 185 | 3 | 0.040 | -225 | 0.20 |
| 30 | 25 | 4 | 0.155 | 180 | 7 | 0.042 | -225 | 0.24 |
| 50 | 25 | 7 | 0.160 | 170 | 15 | 0.060 | -225 | 0.27 |
| 70 | 25 | 10 | 0.170 | 165 | 20 | 0.075 | -225 | 0.30 |

[a] The model is given by equation (1)
[b] The values of the three molecular levels $\varepsilon_1$, $\varepsilon_2$ and $\varepsilon_3$, were experimentally determined from $I$ vs. $V_g$-$V_{sd}$ data.

The energetic difference between the main two energy levels $\Delta\varepsilon_{1,2}$ (= $\varepsilon_1 - \varepsilon_2$) = 160mV (for $V_{sd}$ = 10mV), decreases slightly with increasing $V_{sd}$ (140 meV for $V_{sd}$ = 70 mV). This reduction in the value $\Delta\varepsilon_{1,2}$ has been observed in other molecular tunnel junctions [22], and is associated with the static Stark effect [46]. This shift was taken into consideration in the fitting procedure. The third level, $\varepsilon_3$, responsible for the degeneracy point outside the gate voltage window ($V_g$ ~3.5 V) is included to account for the increase in conduction for $V_g$ < -2 V, but its position with respect to $\varepsilon_1$, *i.e.*, $\Delta\varepsilon_{1,3}$ (= $\varepsilon_1 - \varepsilon_3$), was kept constant in the model as its exact value could not be measured accurately. Highly asymmetric tunneling rates $\gamma_L \neq \gamma_R$ for the main two levels had to be employed to fit the results, with values similar to those reported for other molecular SETs [22]. The reason lies in the fact that these rates present two main effects on the conduction through the junction. On the one hand, the sum of the right and left tunneling rates determines the zero-temperature width of the level at each bias ($\gamma = \gamma_L + \gamma_R$), and therefore the highest rate is fixed by the experimental widths of the transport excitations. On the other hand, the lowest rate limits the current through the junction, and therefore is fixed by the experimentally observed value.

**Discussion**

Figures 2-4 show the excellent agreement between the experimental data and the model, which indicates that the complex temperature dependence of the tunneling current in all three transport regimes arises from the thermal broadening of the Fermi electronic occupation distribution in the leads. The quantitative degree of agreement between experiment and theory is easier to appreciate in Figures 5 which displays the behavior of $I$ as a function of $T$ for two bias voltages ($V_{sd} =$ 10 and 50mV, respectively) with the values of $V_g$ indicated by vertical lines in Figs. 4a and 4c specifically selected to sample the different conduction regimes. These figures clearly show how the electrostatic potential controls the conduction regime across the molecule: *i*) the junction is in the Coulomb blockage regime for $V_g$ = -1.5 (black squares) and +0.9 V (green triangles) for $V_{sd}$ = 10 mV (Fig. 5a) and for $V_g$ = -1.5 and +1.1 V and $V_{sd}$ = 50 mV (Fig. 5b). In this regime $I$ exponentially increases with $T$ (see Supplementary Note 2 and Supplementary Figure 3 for a discussion of a crossover into a temperature-independent transport as the temperature is further decreased in this regime); *ii*) the junction is at a charge

degeneracy point for $V_g$ = -0.3 V and $V_{sd}$ = 10 mV (Fig. 5a) and $V_g$ = -0.2 V and $V_{sd}$ = 50 mV (Fig. 5b). In this regime *I* decreases with *T*; and, finally, *iii*) the junction is at resonance for $V_g$ = -0.7 V and $V_{sd}$ = 10 mV (Fig. 5a) and $V_g$ = -0.8 V and $V_{sd}$ = 50 mV (Fig. 5b). In this regime *I* depends weakly on *T*.

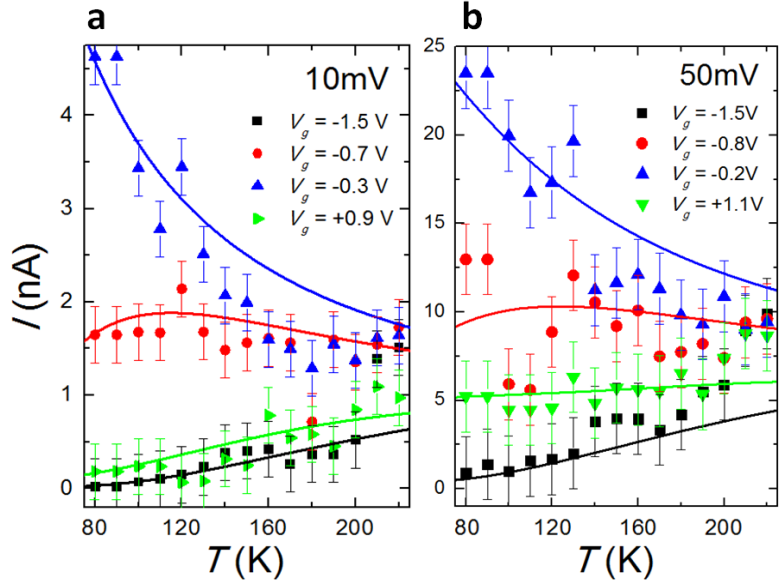

**Figure 5 │ The tunneling current as function of temperature and applied bias. a**, Evolution of the tunnel current with temperature at $V_{sd}$ = 10mV for four different gate voltages ($V_g$ = -1.5, -0.7, -0.3 and +0.9V). **b**, Same for $V_{sd}$ = 50mV ($V_g$ = -1.5, -0.8, -0.2 and +1.1V). The solid lines in both panels represent fittings to the single-level tunneling transport model in equation (1) with the parameters given in Table 1. The error bars represent the uncertainty in the determination of the value of the tunneling current due to the noise of the measurements.

In conclusion, our measurements reveal a complex temperature-dependent behaviour of charge transport phenomena in a ferrocene-based single-molecule junction over a wide range of bias and gate voltages. Exceptionally, we find three transport regimes in which the tunneling current across the molecule increases, decreases, or remains constant as a function of temperature, depending on whether the junctions are tuned into the Coulomb blockade, the charge degeneracy point, or the resonant transport regimes. All of our observations can be rationalized by a Landauer-type single-level model, where the complex temperature-dependent tunneling behavior can be explained by the broadening of the Fermi distributions accounting for the occupation energies of electrons in the transistor leads. The results described in this article represent an important step forward in explaining the temperature-dependent charge transport measurements observed in SAM-based and other single-molecule tunneling junctions and hopefully help to improve existing models of charge transport. Currently, we are investigating junctions with asymmetrically positioned Fc moieties inside the junctions.

**Methods**

**Synthesis of AcS(CH$_2$)$_4$Fc(CH$_2$)$_4$Sac**. We prepared AcS(CH$_2$)$_4$Fc(CH$_2$)$_4$SAc from native ferrocene in three steps using well-established procedures (See Supplementary Methods, Supplementary Fig. 8 and Refs 47 and 48). All compounds were characterized with $^1$H NMR, $^{13}$C NMR, and ESI HRMS. Briefly, ferrocene was substituted with 4-bromobutanoyl chloride via a Friedel-Crafts acylation to yield (Br(CH$_2$)$_3$CO)$_2$Fc (1,5 g, 45% yield). Next, this compound was reduced with borane-tert-butylamine to yield (Br(CH$_2$)$_4$)$_2$Fc (1.2 g, 88% yield). The bromine functionality was converted to a thioacetate functionality with potassium thioacetate and the AcS(CH$_2$)$_4$Fc(CH$_2$)$_4$SAc was isolated with a yield of 95% (1.0 g).

This thioacetate derivative is stable and was stored under an atmosphere of N$_2$ at -20 °C. We found that the isolation of the free dithiol was problematic because of disulfide formation. Hence our choice to isolate and use the thioacetate protected derivative which was deprotected in situ for self-assembly of the junction.

**Device Fabrication**. For the fabrication of the SETs thin gold nanowires (15 nm thick, 50 nm wide) patterned via electron-beam lithography were deposited on top of an Al/Al$_2$O$_3$ back gate electrode (1-2nm-thick oxide barrier) [40]. We used a feedback-controlled electromigration-induced breaking procedure to narrow the nanowires. Once the current dropped close to the universal single-channel conductance ($G_0$), the current flow was stopped at which point the devices were left untouched for two hours during which the gold atoms reorganize due to surface tension to finally create stable gaps of 1-2 nm. After the self-rupture of the wires, the chip was immersed in a 1 mM solution of AcS(CH$_2$)$_4$Fc(CH$_2$)$_4$SAc in 20 mL toluene and 5 mL methanol along with 10 mg of n-Bu$_4$NCN (CAS 10442-39-4). Here, n-Bu$_4$NCN in this toluene/methanol mixture deprotects the thioacetate mildly [49,50]. Finally, the devices were mounted in a low temperature probe and cooled down to liquid nitrogen temperatures. Out of 322 devices, 11 (~3.5%) showed molecular transport characteristics, but only 2 junctions (~0.6%) were stable enough for studies as a function of temperature (80-220K shown in the main text, and 80-250K shown in the Supplementary Note 1 and Supplementary Figure 1), with molecular conformational changes as the most possible cause for device instability in the other cases. We note that charge transport measurements of molecular SETs at elevated temperatures are challenging because of the high chances for the molecule to change its conformation. We also want to remark that the yield in our experiments is similar to those in most reports of molecular single-electron transistors (see, *e.g.*, Refs. [22,45] , which are most closely related to this work). To ensure that our devices contained a ferrocene-based molecule, we include in the Supplemental Information details of the junctions fabrication and characterization of the nano-gaps (Supplementary Methods: 'Electromigration- breaking of nanowires and device statistics' and Supplementary Figure 5), detailed information on the protocols for recognition and selection of ferrocene-based junctions, including measurements with magnetic fields dependent transport (Supplementary Methods: 'Identification of single-molecule SETs for this study' Supplementary Figure 6), and tests of the electrical characteristics of empty junctions as a function of temperature (Supplementary Methods: 'Three-terminal SET circuit characterization and its temperature behavior' and Supplementary Figure 7).

**Transport measurements**. Transport measurements were performed in DC by applying the bias voltage and measuring the current via a Keithley 6430 sub-femtoamp remote sourcemeter, while a Keithley 2400 was used to gate the transistor. The device was mounted on the sample holder of a home-made measuring probe with temperature control immersed in a liquid nitrogen dewar.

**References**

[1] Nijhuis, C. A., Reus, W. F., Barber, J., Dickey, M. D. & Whitesides, G. M. Charge transport and rectification in arrays of SAM-based tunneling junctions. *Nano Lett.* **10,** 3611–3619 (2010).

[2] Yuan, L., Breuer, R., Jiang, L., Schmittel, M. & Nijhuis, C. A. A molecular diode with a statistically robust rectification ratio of three orders of magnitude. *Nano Lett*. **15,** 5506–5512 (2015).

[3] Yuan, L. *el al*. Controlling the direction of rectification in a molecular diode. *Nat. Commun*. **6,** 6324 (2015).

[4] Yoon, H. J. *et al*. Rectification in tunneling junctions: 2,2′-bipyridyl-terminated n-alkanethiolates. *J. Am. Chem. Soc*. **136,** 17155–17162 (2014).

[5] Ho Choi, S., Kim, B. & Frisbie, C. D. Electrical resistance of long conjugated molecular wires. *Science*. **320,** 1482 (2008).

[6] Yan, H. *et al*. Activationless charge transport across 4.5 to 22 nm in molecular electronic junctions. *Proc. Natl. Acad. Sci. U.S.A*. **110,** 5326 (2013).

[7] Smith, C. E. *et al*. Length dependent nanotransport and charge hopping bottlenecks in long thiophene-containing π-conjugated molecular wires. *J. Am. Chem. Soc*. **137,** 15732–15741 (2015).

[8] Lafferentz, L. *et al.* Conductance of a single conjugated polymer as a continuous function of its length. *Science* **323,** 1193–1197 (2009).

[9] Koch, M., Ample, F., Joachim, C. & Grill, L. Voltage-dependent conductance of a single graphene nanoribbon. *Nature Nanotech*. **7,** 713–717 (2012)

[10] Livshits, G. I. *et al.* Long-range charge transport in single G-quadruplex DNA molecules. *Nature Nanotech.* **9,** 1040–1046 (2014).

[11] Li, W. *et al.* Temperature and force dependence of nanoscale electron transport via the Cu protein azurin. *ACS Nano* **6,** 10816–10824 (2012).

[12] Sepunaru, L., Pecht, I., Sheves, M. & Cahen, D. Solid-state electron transport across azurin: from a temperature-independent to a temperature-activated mechanism. *J. Am. Chem. Soc*. **133,** 2421–2423 (2011).

[13] Kumar, K. S., Pasula, R. R., Lim, S. & Nijhuis, C. A. Long-range tunneling processes across ferritin-based junctions. *Adv. Mater*. 10.1002/adma.201504402 (2015).

[14] Joachim, C. & Ratner, M. A. Molecular electronics: some views on transport junctions and beyond. *Proc. Natl. Acad. Sci. U S A*. **102,** 8801 (2005).

[15] Marcus, R. A. On the theory of oxidation-reduction reactions involving electron transfer I. *J. Chem. Phys.* **24,** 966 (1956).

[16] McCreery, R. L. Molecular electronic junctions. *Chem. Mater.* **16,** 4477–4496 (2004).

[17] Qi, Y. *et al*. Filled and empty states of alkanethiol monolayer on Au (1 1 1): fermi level asymmetry and implications for electron transport. *Chem. Phys. Lett*. **511,** 344 (2011).

[18] Migliore, A., Schiff, P. & Nitzan, A. On the relationship between molecular state and single electron pictures in simple electrochemical junctions. *Phys. Chem. Chem. Phys*. **14,** 13746–13753 (2012).

[19] Moth-Poulson, K. & Bjørnholm, T. Molecular electronics with single molecules in solid-state devices. *Nature Nanotech*. **4,** 551–556 (2009).

[20] Heimel, G. *et. al*. Charged and metallic molecular monolayers through surface-induced aromatic stabilization. *Nature Chemistry* **5,** 187–194 (2013).

[21] Perrin, M. L. *et al*. Large tunable image-charge effects in single-molecule junctions. *Nature Nanotech.* **8,** 282–287 (2013).

[22] Poot, M. *et al*. Temperature dependence of three-terminal molecular junctions with sulfur end-functionalized tercyclohexylidenes. *Nano Lett*. **6,** 1031–1035 (2006).

[23] Bergren, A. J., McCreery, R. L., Stoyanov, S. R., Gusarov, S. & Kovalenko, A. Electronic characteristics and charge transport mechanisms for large area aromatic molecular junctions. *J. Phys. Chem. C* **114,** 15806–15815 (2010).

[24] Diéz-Pérez, I. *et al*. Rectification and stability of a single molecular diode with controlled orientation. *Nat. Chem*. **1,** 635–641 (2009).

[25] Venkataraman, L., Klare, J. E., Nuckolls, C., Hybertsen, M. S. & Steigerwald, M. L. Dependence of single-molecule junction conductance on molecular conformation. *Nature* **442,** 904–907 (2006).

[26] Chiechi, R. C., Weiss, E. A., Dickey, M. D. & Whitesides, G. M. Eutectic gallium–indium (EGaIn): a moldable liquid metal for electrical characterization of self-assembled monolayers. *Angew. Chem. Int. Ed*. **47,** 142–144 (2008).

[27] Fracasso, D., Valkenier, H., Hummelen, J. C., Solomon, G. & Chiechi, R. C. Evidence for quantum interference in SAMs of arylethynylene thiolates in tunneling junctions with eutectic Ga-In (EGaIn) top-contacts. *J. Am. Chem. Soc*. **133,** 9556–9563 (2011).

[28] McCreery, R. L. & Bergren, A. Progress with molecular electronic junctions: meeting experimental challenges in design and fabrication. *Adv. Mater*. **21,** 4303–4322 (2009).

[29] Haj-Yahia, A.-E. *et al*. Substituent variation drives metal/monolayer/semiconductor junctions from strongly rectifying to ohmic behavior. *Adv. Mater.* **25,** 702–706 (2013).

[30] Bowers, C. M. *et al*. Characterizing the metal–SAM interface in tunneling junctions. *ACS Nano* **9,** 1471–1477 (2015).

[31] Park, H. *et al*. Nanomechanical oscillations in a single-C60 transistor. *Nature* **407,** 57–60 (2000).

[32] Park, J. *et al*. Coulomb blockade and the kondo effect in single-atom transistors. *Nature* **417,** 722–725 (2002).

[33] Liang, W., Shores, M. P., Bockrath, M., Long, J. R. & Park, H. Kondo resonance in a single-molecule transistor. *Nature* **417,** 725–729 (2002).

[34] Kubatkin, S. *et al*. Single-electron transistor of a single organic molecule with access to several redox states. *Nature* **425,** 698–701 (2003).

[35] Haque, F., Langhirt, M., del Barco, E., Taguchi, T. & Christou, G. Magnetic field dependent transport through a Mn4 single-molecule magnet. *J. Appl. Phys*. **109,** 07B112 (2011).

[36] Song, H. *et al*. Observation of molecular orbital gating. *Nature* **462,** 1039–1043 (2009)

[37] Jeong, H. *et al*. Redox-induced asymmetric electrical characteristics of ferrocene-alkanethiolate molecular devices on rigid and flexible substrates. *Adv. Funct. Mater*. **24,** 2472–2480 (2014).

[38] Müller-Meskamp, L. *et al*. Field-emission resonances at tip/α,ω-mercaptoalkyl ferrocene/Au interfaces studied by STM. *Small* **5,** 496–502 (2009).

[39] Mentovich, E. D. *et al*. Gated-controlled rectification of a self-assembled monolayer-based transistor. *J. Phys. Chem. C* **117,** 8468–8474 (2013).

[40] Nijhuis, C. A., Reus, W. F., Siegel, A. C. & Whitesides, G. M. A molecular half-wave rectifier. *J. Am. Chem. Soc*. **133,** 15397–15411 (2011).

[41] Henderson, J. J., Ramsey, C. M., del Barco, E., Mishra, A. & Christou, G. Fabrication of nano-gapped single-electron transistors for transport studies of individual single-molecule magnets. *J. Appl. Phys*. **101,** 09E102 (2007).

[42] Datta, S. Electrical resistance: an atomistic view. *Nanotechnology* **15,** S433-S451 (2014).

[43] Datta, S. *Lessons from Nanoelectronics: A New Perspective on Transport* Vol. 1 (World Scientific, Singapore, 2012).

[44] Jauho, A. P., Wingreen, N. S. & Meir, Y. Time-dependent transport in interacting and noninter-acting resonant-tunneling systems. Phys. Rev. B 50, 5528-5544 (1994).

[45] de Leon, N. P., Liang, W., Gu, Q. & Park, H. Vibrational excitation in single-molecule transistors: deviation from the simple Franck−Condon prediction. *Nano Lett*. **8,** 2963-2967 (2008).

[46] Xue, Y. & Ratner, M. A. Microscopic study of electrical transport through individual molecules with metallic contacts. I. Band line-up, voltage drop, and high-field transport. *Phys. Rev. B*. **68,** 115406 (2003).

[47] Vulugundam, G., Kumar, K., Kondaiah, P. & Bhattacharya, S. Efficacious redox-responsive gene delivery in serum by ferrocenylated monomeric and dimeric cationic cholesterols. *Org. Biomol. Chem*. **13,** 4310-4320 (2015).

[48] Gharib, B. & Hirsch, A. Synthesis and characterization of new ferrocene-containing ionic liquids. *Eur. J. Org. Chem*. **19,** 4123-4136 (2014).

[49] Holmes, B. T. & Snow, A. W. Aliphatic thioacetate deprotection using catalytic

tetrabutylammonium cyanide. *Tetrahedron* **61,** 12339–12342 (2005).

[50] Park, T. *et al.* Formation and structure of self-assembled monolayers of octylthioacetates on Au(111) in catalytic tetrabutylammonium cyanide solution. *Bull. Korean Chem. Soc.* **30,** 441-444 (2009).

# Supplementary Information: Electrostatic Control over Temperature-Dependent Tunneling across a Single Molecule Junction

*Alvar R. Garrigues[1], Lejia Wang[2], Enrique del Barco[1]\* & Christian A. Nijhuis[2,3]\**

[1] Department of Physics, University of Central Florida, Orlando, Florida 32816 - USA

[2] Department of Chemistry, National University of Singapore, 3 Science Drive 3, Singapore 117543

[3] Graphene Research Centre, National University of Singapore, 6 Science Drive 2, Singapore 117546

(\*) CORRESPONDING AUTHORS: E.d.B: Tel.: (+1) 823 0755, e-mail: delbarco@ucf.edu; C.A.N: Tel.: (+65) 6516 2667, e-mail: christian.nijhuis@nus.edu.sg

**1. SET results on second molecule**

Figure S1a shows a 3D plot of $I$ vs. $V_g$ as a function of temperature for $V_{sd} = 10$ mV in a second ferrocene-based molecule measured in this study. The data were collected as the gate voltage was continuously swept from -2.5 to +2.5 V at each temperature ranging from 80 to 250 K. Similarly to the single molecule SET described in the main text, the results allow following the behavior of the current when changing the temperature within the full range of gate voltages, enabling the analysis of the temperature dependence of the current through the junction in different transport regimes. Fig. S1a shows the experimental data and Fig. S1B shows the modelling results. Similar to the device described in the main text, the results show: *i*) the Coulomb blockade regime (regime 1), with $I$ increasing exponentially with temperature (follow blue arrow at $V_g = 2.5$ V in the S1a data and its projection in the lower $V_g$-$T$ panel in Fig. S1a); *ii*) the degeneracy points (regime 2), with *I* decreasing markedly with *T* (red arrows); and, *iii*) intermediate zones (regime 3, with just a slight variation of *I* with *T* (black arrow) when the molecular level matches the electrostatic potential of one of the leads (resonance).

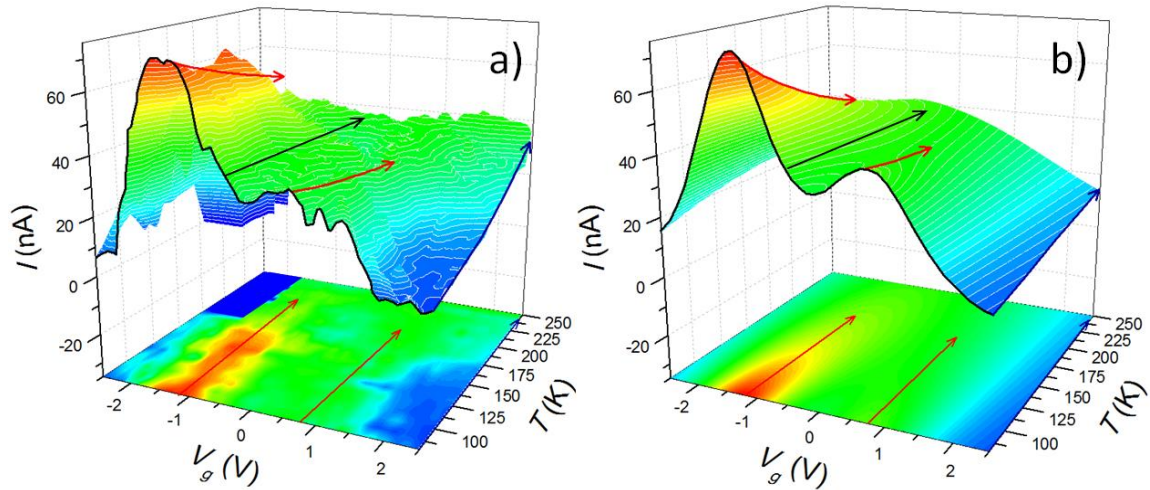

**Figure S1: a)** 3D plot of the evolution of the tunnel current through a second S-$(CH_2)_4$-Fc-$(CH_2)_4$-S junction vs. gate voltage as the temperature is increased from 80 to 250 K and with an applied bias voltage of 10 mV. The evolution of the two charge points ($V_g = -1.25$ V and 0.8 V), whose magnitude decreases with increasing temperature are indicated with red arrows in both the 3D data and the corresponding 2D horizontal projection in the $V_g$-$T$ plane. Similarly, the increase of the current in the Coulomb blockade area is indicated with the blue arrow. **b)** Corresponding response of the junction as calculated from the single-level tunneling transport model in Eqn. (1) in the main text using the parameters given in *Table S1*.

**Table S1:** Parameters employed in fitting the experimental data in Figures S1 and S2 to the single-level model ( Eq. 1)[a].

| $V_b$ (mV) | $\varepsilon_1$ (mV) | $\gamma_{1L}$ (mV) | $\gamma_{1R}$ (mV) | $\varepsilon_2$ (mV) | $\gamma_{2L}$ (mV) | $\gamma_{2R}$ (mV) |
|---|---|---|---|---|---|---|
| 10 | 62.5 | 30 | 3 | 165 | 45 | 1.9 |

[a] Two molecular levels with energies $\varepsilon_1$ and $\varepsilon_2$ with respect to the Fermi energy of the electrodes at zero bias were determined from the experimental data.

The results for the second molecule were fitted to the model given by Eqn. (1) in the main text using the parameters listed in Table S1. Figure S1b shows a 3D plot of the calculated current through the junction. Two molecular levels were employed to account for the two degeneracy points observed within the window of electrical potentials. Again, only two free-fitting parameters were used per molecular level to fit the results with Eqn. (1), *i.e.*, the respective tunneling rates ($\gamma_{L,R}$) associated to the flow of electrons from the leads into/out of the molecule. Here, the values of $\gamma_{L,R}$ were ~10 times larger than those used to fit the experimental data discussed in the main text; these large values of $\gamma_{L,R}$ are in agreement with the larger current magnitude (~10 times larger) and excitations widths observed in the second molecule, and we associate them to a stronger connection of the molecule with the transistor leads.

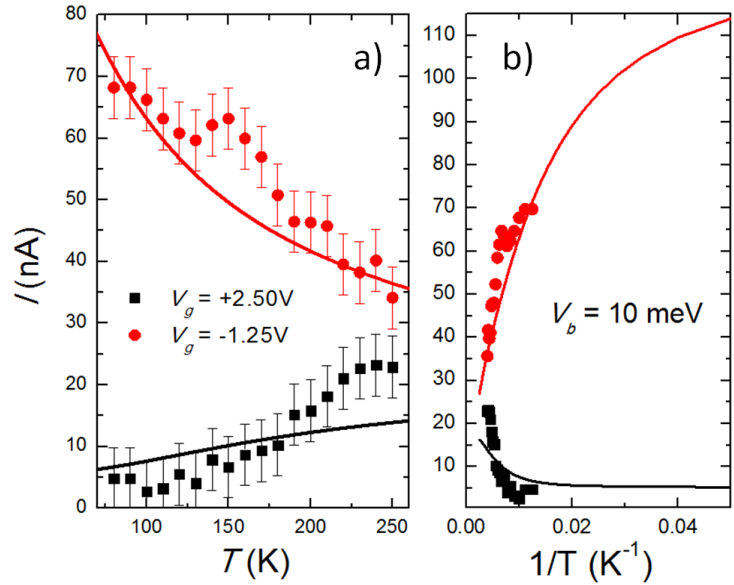

**Figure S2: a)** Experimental and calculated evolution of the tunnel current with temperature at 10 mV-bias for two different gate voltages, $V_g$ = -1.25 V (charge degeneracy point) and +2.5 V (Coulomb blockade regime). **b)** The same results shown as a function of the inverse temperature in an extended range of temperature to visualize the transition between the temperature-independent and –dependent conduction regimes, which for this molecule occurs at ~100 K.

Figures S2a and S2b display the behavior of *I* as a function of *T* for two values of $V_g$ specifically selected to sample the coulomb blockade regime, $V_g$ = 2.5 V (black squares), and the charge degeneracy point, $V_g$ = -1.25 V (red circles), for $V_b$ = 10 mV. With the same overall behaviors as those found for the molecule discussed in the main text. Specifically, with *I* increasing with *T* in the Coulomb blockade regime, and *I* decreasing with *T* at the charge degeneracy point. The transition between the temperature-dependent and temperature-independent behavior of the current in the Coulomb blockade regime can be clearly seen in Fig. S2b (black circles) to occur at about 100 K for this molecule. The continuous lines in Figure S2 represent the fittings to the model, in good overall agreement with the data, with the exception of the data at the highest temperatures. This is likely due to a molecular conformational change (e.g. a slight displacement of the molecule with respect to the SET leads) or a bond-fluctuation due to a change in the molecular attachment to the electrodes [1] occurring at ~175 K which shifted the charge degeneracy points ~1V towards positive gate voltages, also affecting the overall conduction through the junction for higher temperatures. To correct for this displacement, the data above ~175 K was correspondingly shifted to higher gate voltages, as can be observed in the horizontal projection of the data in Fig. S1a, with no data in the $T > 175K - V_g < -1.5V$ area (dark blue area).

## 2. Crossover between temperature dependent and independent transport in the Coulomb blockade regime

Our junctions present clear evidences of the transition between the thermally assisted and the coherent tunneling regimes in the Coulomb blockade regime (*e.g.*, $V_g = -1.5$ V in Fig. 5a in the main text, or ), with the data/calculation bending as the temperature is decreased close to 80 K. Figure S3 extends the calculations presented in the main text into lower temperatures to give a better idea of the crossover between the temperature-dependent and -independent conduction regimes. This transition depends strongly on the applied gate and bias voltages. Figure S3a shows the case for $V_b = 10$ mV, where the crossover temperature in the Coulomb blockade regime ($V_g$ = -1.5 V) lies at ~65 K, which is slightly lower than the lowest temperature used in the experiments. Figure S3b shows a log-plot of the calculated current for $V_b$ = 10 mV and the behavior of the crossover temperature with the applied gate voltage; this crossover linearly increases with bias.

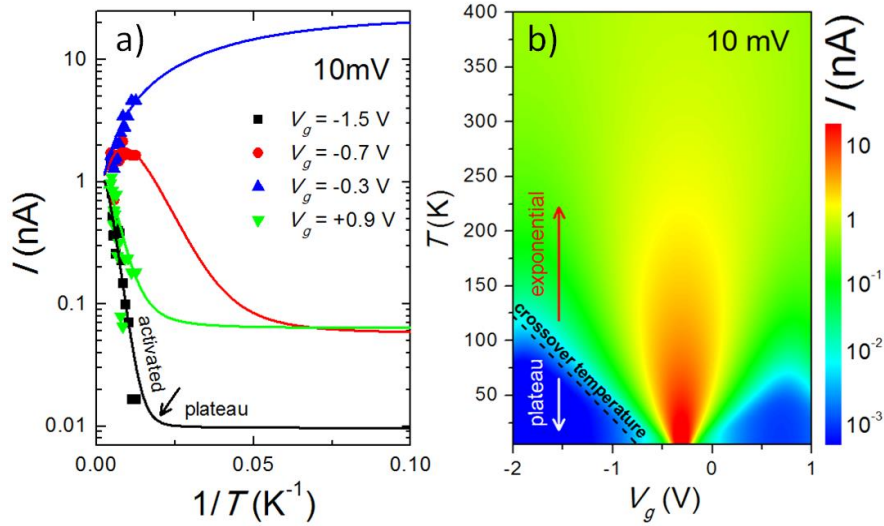

**Figure S3: a)** Experimental (symbols) and calculated (lines) current versus the inverse of temperature for four different gate voltages ($V_g$ = -1.5, -0.7, -0.3 and +0.9 V) at $V_b$ = 10 mV. **b)** Contour color-code plot of the temperature behavior of the current (in log-scale) with respect to gate voltage for $V_b$ = 10 mV, illustrating how the crossover between the thermally-assisted and temperature-independent regimes varies with the applied gate.

## 3. Reversibility of the temperature behavior of the tunnel current through the SET

Figure S4 shows measurements of the tunneling current through the molecular transistor described in the main text (molecule 1) obtained at different times between which the temperature of the sample had been raised to ~200K. Specifically, Fig. S4a contains the same data shown in Fig. 2a (main text), corresponding to the preliminary characterization of the Coulomb blockade response of the molecular SET at ~80K, with the two charge degeneracy points at $V_g$ = 0.25 and 2.5V ($V_{sd}$ = 0), and an energy separation between the corresponding charge states of $\Delta\varepsilon$ ~110meV. After that preliminary characterization, the temperature was raised to ~200K, while taking some sample $I_{sd}$-$V_g$ curves at $V_b$ = 10mV at a few intermediate temperatures as a quick check of the temperature behavior of the device and to test its stability (procedure repeated with all detected molecules). After that, the temperature was decreased back to ~80K. Fig. S4b shows the $I_{sd}$-$V_g$ curve at $V_b$ = 10mV obtained after that (same data in Fig. 2b in the main text). Only a small change in the position of the degeneracy points (now at $V_g$ = -0.3 and 1.7V), and a change in energy difference between the two charge states (now Δε ~140meV) was observed, while the overall behavior of the molecular SET was unaltered. As discussed in Section 1, this may be associated to small conformational changes of the molecule within the transistor leads or bond-fluctuations due to alterations in the attachment of the molecule to the respective electrodes [1]. Subsequently, the temperature behavior of the device was studied while slowly increasing the temperature up to 220K, from which the central results of this work where obtained. At the conclusion of the study, the temperature was decreased again back to ~80K and a new Coulomb blockade contour plot measured (Fig. S4c), which shows that the molecule behaved similarly after the thermal cycling (with $V_g$ = -0.4 and 2.2V, and Δε ~140meV), demonstrating the reversibility of its temperature behavior.

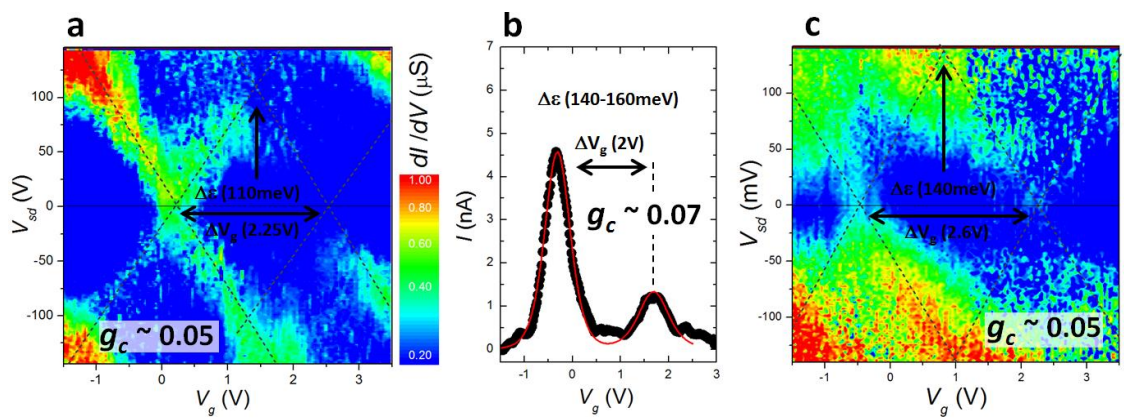

**Figure S4: a)** Differential conductance of a junction with S-$(CH_2)_4$-Fc-$(CH_2)_4$-S at $T$ = 80 K (same data as in Fig. 2a in the main text). The color code represents the conductance ($dI/dV$) through the SET as a function of $V_{sd}$ and $V_g$. **b)** Current vs. gate voltage for $V_{sd}$ = 10mV at $T$ = 80K obtained after the first thermal cycling (same data than in Fig. 2b in the main text). **c)** Differential conductance of the same junction at $T$ = 80 K after the temperature dependency study had been completed (using the same color code scale than in **a** for the conductance).

## 4. Characterization of the molecular SETs

*4.1. Electromigration-breaking of nanowires and device statistics*

For the fabrication of the nano-gap SETs we follow a recipe that we have refined over recent years by which gaps separating the source and drain transistor leads of the order of 1-2nm are achieved in high yields (>80%) at room temperature, which is similar to that reported by van der Zant and collaborators [2]. For this, a feedback-controlled electromigration breaking procedure (Fig. S5a) is employed first to narrow the wires down until obtaining resistances ranging from 1 to 100kΩ in most of the wires, 75% (*i.e.*, the wire is not completely broken). Figure S5b shows the statistics corresponding to the wires prepared for the studies presented in this article (with a sample of over 300 nanowires).

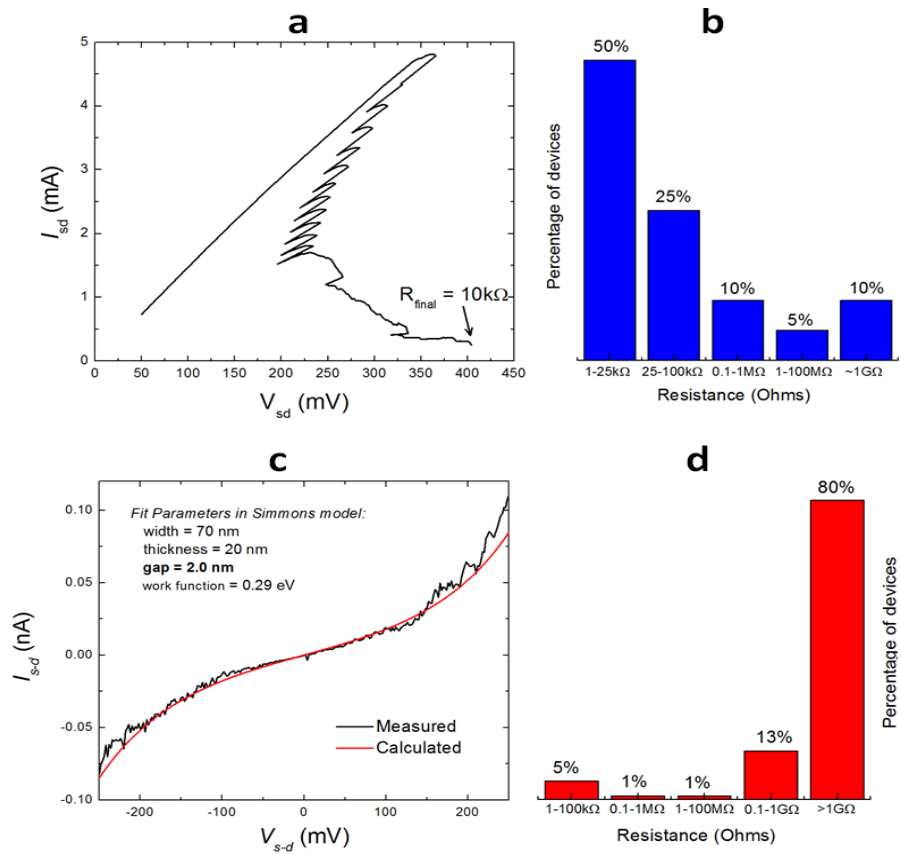

**Figure S5: a)** Electron-migrating narrowing of a Au nanowire. **b)** Statistics showing the resistance of the resulting constricted wires. **c)** Tunneling current through a 2nm Au nanogap, as determined by fitting to the Simmons model. **d)** Statistics of the resulting nanogaps.

After the narrowing of the nanowires, the chips are left untouched for 1-2 hours, during which time the reorganization of the gold surface leads to the eventual rupture of the wires, generating gaps of 1-2nm in most of the wires. Figures S5c and S5d present the characteristic tunneling current of the nano-gapped wires and the statistics on the batch used for this study, respectively. In the case of this study, 93% of the resulting wires present resistances within the vicinity of 1GΩ, indicative of 1-2nm gaps. In the original report on this procedure by van der Zant and collaborators the authors state that *no evidence of formation of gold nanoparticles in the gaps was found out of 300 junctions prepared with this recipe* [2].

After the gaps have been formed, a solution with the molecules under study is used to bridge the nanogaps during a 1-hour deposition process in with the solution is put in direct contact with the chip of SETs. After that, the chip is inserted into the low-temperature probe and brought to base temperature, where it is tested for variations in the gaps' resistances (decrease) indicative of the presence of a molecule bridging the gap separating the source and drain electrodes. As mentioned in the main text, out of 322 devices, 11 (~3.5%) showed molecular transport characteristics (see section 4.2 below for details). Note that this procedure prevents having to break the Au nanowires at low temperature, for which the molecules had to be deposited in advance. Allowing the room-temperature gold surface reorganization to complete the nanowire rupture helps in obtaining small gaps with high mechanical stability and much higher yields (~90%) than low-T breaking (with past yields limited to 50% in our group). In addition, since the rupture is performed at room temperature, the molecules can be deposited after the gap has been formed, minimizing substantially (almost eliminating) the possibility of formations of metallic islands that can be mistaken by molecules or thermal degradation of the molecule during electron migration. It is for this important reason that signatures of Coulomb blockade with charge states separated by 100meV or more are taken as signatures of molecular presence, as discussed below.

*4.2. Identification of single-molecule SETs for this study*

As a test routine after deposition of molecules in the nano-gapped SETs, the chip is cooled down to $T = 80K$ and the tunnel current $I_{sd}$ is tested at each transistor by applying a small bias voltage ($V_{sd} = 10mV$). A decrease of an order of magnitude or more in the tunneling current through a transistor with respect to its value before the deposition of molecules is taken as an indication of an incidence in the nano-gap.

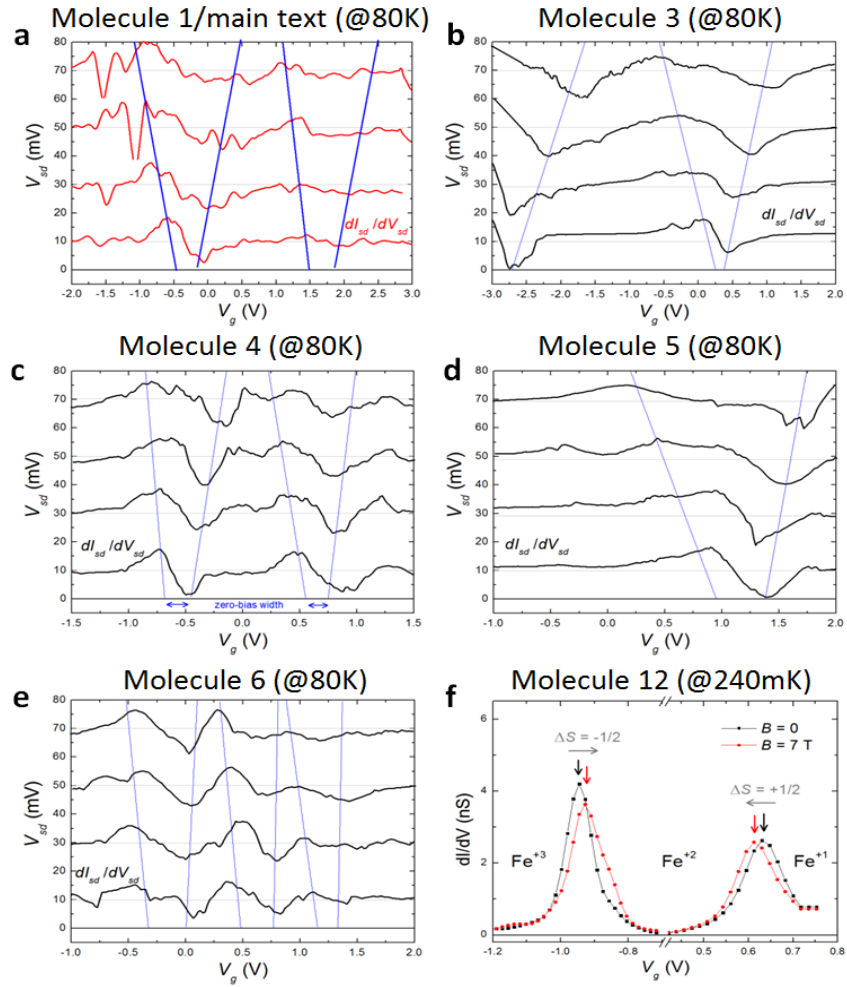

**Figure S6: a)** Differential conductance as a function of gate voltage for several bias voltages (10, 30, 50 and 70mV) recorded at 80K for the molecule described in the main text of this article (molecule 1), corresponding to the derivative ($dI_{sd}/dV_{sd}$) of the data in Figure 2b in the main text. **b-e)** Same measurements for molecules 3 to 6. **f)** $dI_{sd}/dV_{sd}$ vs $V_g$ measured at $T = 4K$ and $V_{sd} = 10mV$ with and without a 7 Tesla magnetic field applied for another molecule.

For those transistors, the differential conductance is monitored as a function of the gate voltage for four different source-drain voltages (*i.e.*, $V_{sd}$ = 10, 30, 50 and 50mV). Figure S6a-e shows examples of five out of the eleven molecules recognized during the studies performed at 80K, including the molecule discussed in the main text (molecule 1), which is displayed in Fig. S6a. As clearly observed in Figures S6a-e, charge transport excitations are observed at different gate voltages, whose width (separation between peaks and dips in the $dI_{sd}/dV_{sd}$) increases as the bias voltage ($V_{sd}$) is increased (follow blue lines in Figs. S6a-e). The linear dependence of the transport excitations with the bias voltage is indicative of the characteristic response of a molecular SET, with charge degeneracy points separating different molecular charge states, *i.e.*, areas with suppression of tunneling current by the Coulomb blockade effect. Identification of charge states separated by $\Delta\varepsilon$ = 100meV or more (bias voltage values for which adjacent blue lines would meet), together with the fact that the gaps are formed at room temperature in the absence of molecules, is taken as proof of a molecular presence in these eleven SET junctions. Note that strong renormalization effects are known to lower significantly (one order of magnitude) the energy separation between molecular charge states as a result of the presence of the metallic electrodes, which act as electrical mirrors. Forty nine other junctions showing gate and bias dependencies but not meeting the criteria above (*i.e.*, $\Delta\varepsilon$ < 100meV, non-linear dependence or low-resistance in the junction before the deposition of molecules) were disregarded for further study.

It is known that the variable electrostatic environment between the electrodes in the SETs makes molecules in general (and ferrocene in particular) to display different charge states after deposition [3]. Indeed, the charge state of the molecule can be tested by monitoring the relative shift in gate voltage of the charge degeneracy points under the action of an applied magnetic field, $\Delta V_c = -2g\mu_B C_T/C_g \Delta S$, where g is the electron

Lande factor, $\mu_B$ is the Bohr magneton, $C_T$ is the total capacitance of the junction, $C_g$ is the gate capacitance, and $\Delta S$ is the spin change [4]. To check on this, a chip of forty SETs prepared according to this protocol was mounted in a low-T probe and inserted in a dewar with a superconducting magnet and brought down to sub-milliKelvin temperatures for studies in the presence of a magnetic field. Fig. S6f shows the measured tunnel current through one of the chip's junction at 240mK (black data), which clearly displays two charge degeneracy points at $V_g$ = -0.95 and +0.63 V. Note that the signal is substantially less noisy than that from measurements at 80K (molecular junctions are also more stable at low temperature). Upon application of a magnetic field of 7 Tesla, the peaks move towards different directions in gate voltage, with the first/second peak shifting to higher/lower voltages, respectively. The corresponding analysis determines a change of spin $\Delta S$ = -1/2 for the first degeneracy point, while crossing the second degeneracy point involves a positive change of spin $\Delta S$ = +1/2. The observed spin transitions agree well with the two transport degeneracies separating the following charge states in the ferrocene, $Fe^{3+}$ ($S$ = 1/2) → $Fe^{2+}$ ($S$ = 0) → $Fe^{1+}$ ($S$ = 1/2), and would be difficult to reconcile with a non-molecular origin. This provides further support to the association made in this work between the transport response of the transistors and the ferrocene-based molecules used to construct them.

## *4.3. Three-terminal SET circuit characterization and its temperature behavior*

Once the characteristic molecular signatures, as those shown in Figs. S6a-e, are identified the corresponding molecular junctions were subjected to measurements at a few temperatures between 80 and ~200 K to check the thermal stability of the junctions. As mentioned in the main text, only two molecules survived the first change in temperature (molecule 1 in main text, molecule 2 in *Section 1* of this document), and remained stable for a detailed study within the 80-220K (1) and 80-260K (2) *T*-ranges.

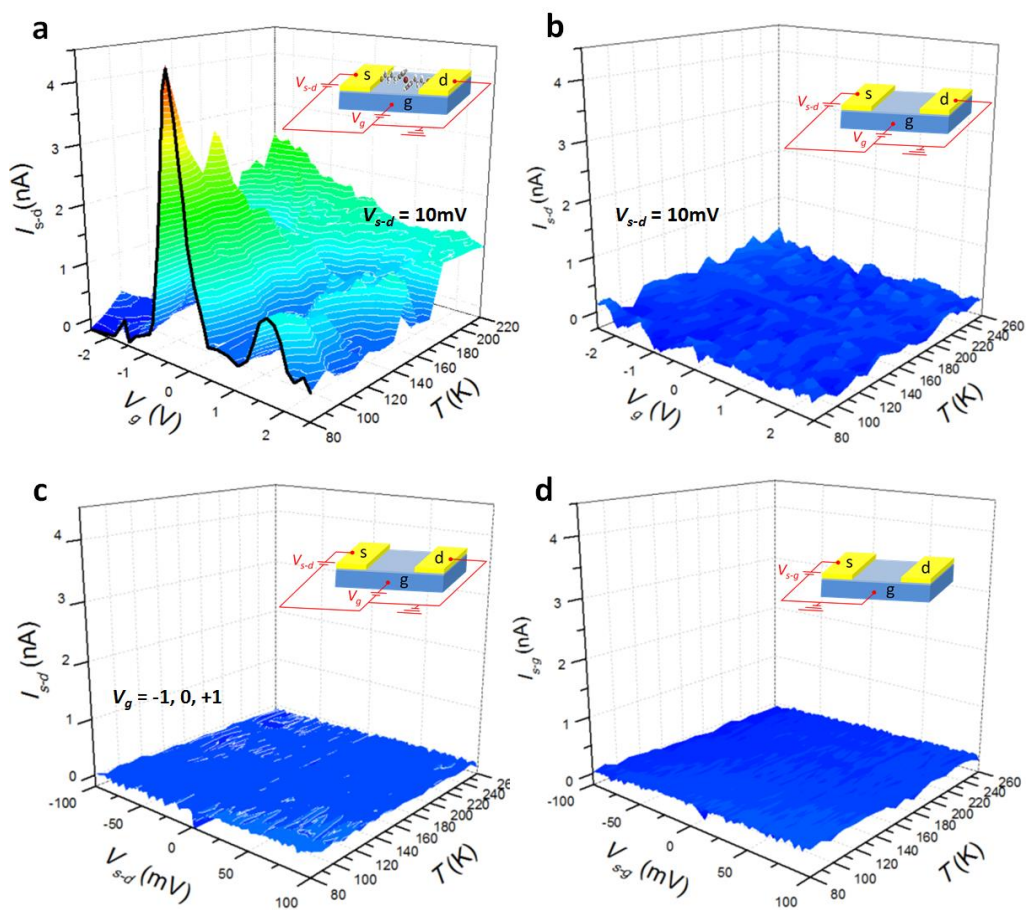

**Figure S7: a)** 3D plot of the evolution of the tunnel current through a S-$(CH_2)_4$-Fc-$(CH_2)_4$-S junction vs. gate voltage as the temperature is increased from 80 to 220 K for $V_{sd}$ = 10 mV. **b)** The same exact measurement as in (a) with an empty junction (no molecule). **c)** Tunnel current through an empty junction as a function of the bias voltage $V_{sd}$ and temperature. The result does not change appreciably in this scale for three different gate voltages $V_g$ = -1, 0, +1. **d)** Tunnel current between the source electrode and the gate as a function of source-gate voltage $V_{s-g}$ and temperature. The same result is obtained between the drain electrode and the gate.

It is important nonetheless to assure that the observed temperature behavior is originated by the molecule in the junction and not by device artifacts. To check this end, the behaviors of several empty junctions (*i.e.*, SETs from the same fabrication batch but without molecules deposited) were examined upon changes in temperature. The results of one empty junction are shown in Figure S7 (similar results were observed in the other junctions examined). For comparison, Fig. 7a shows the contour plot of the tunnel current $I_{sd}$ as a function of gate voltage and temperature for a bias voltage $V_{sd}$ = 10mV (same data as in Fig. 3a in the main text). It is very illustrative to compare Fig. S7a with S7b, since they correspond to the same exact measurement in the molecular junction (S7a) and in the empty junction (S7b). At all gates and temperatures (in the case of the empty junction up to 260K) the current through the empty junction remains negligible ($I_{sd}$ < 0.1nA in all examined empty junctions) when compared to the current through the molecular device (almost two orders of magnitude larger at the charge degeneracy point, and more at higher bias voltages). Note that although the current magnitude in a molecular junction depends greatly on its coupling to the electrodes (see, *e.g.*, the much larger current amplitude in molecule 2 in *Section 1* of this document), it is always expected to be substantially larger than that flowing across an empty gap. But, most importantly, the absence of prominent features in the temperature and gate voltage responses of the empty junctions provides the distinct behavior observed in the molecular junctions a strong proof of their molecular origin. Figure S7c shows the evolution of $I_{sd}$ vs. $V_{sd}$ with temperature in an empty junction (the corresponding tunneling curve and its fit to the Simmons model for $T$ = 80K was shown in Fig. S4c). Again, no appreciable changes are observed within the range of temperatures measured.

Finally, the devices are always checked for leaks between the source/drain transistor leads and the gate electrode. The isolation between them is achieved by a thin (1-2nm) $Al_2O_3$ layer which is fragile and sensitive to drastic electrostatic changes. These barriers are routinely characterized in our laboratories any time a new fabrication batch of SETs is obtained. Voltages above 4.5-5 volts are necessary to break the barrier, and the resulting leaks are easily identifiable during the SET measurements (sometimes the rupture occurs while measuring due to an electrostatic discharge). However, it is important to verify that the small tunneling current between source/drain leads and the gate is not compromised when varying temperature. To evaluate this, the tunnel current between the source and gate electrodes in several empty junctions has been measured within the same temperature range described above and the results in one of the devices displayed in Fig. S7d (identical results were found in the other devices examined). Basically, the tunneling current is again negligible (<0.1nA) when compared to the current through the molecular junctions reported in this article and does not display any appreciable change with increasing temperature.

## 5. Synthesis detail

*General Procedures.* The following chemicals were purchased from Sigma-Aldrich and used without further purification: 4-bromobutanoyl chloride, borane-tert-butylamine complex, ferrocene, anhydrous aluminum chloride and potassium ethanethioate. Solvents for chemical synthesis were freshly distilled prior to use: dichloromethane (DCM) was distilled from calcium chloride and tetrahydrofuran (THF) was distilled from sodium/benzophenone. Deionized water (18.2 MΩ cm) was generated from a water purifier (Purelab Option). All moisture sensitive reactions were performed under a $N_2$-atmosphere. Thin layer chromatography (TLC) glass plates coated with 0.25 mm thick layer of silica gel 60 and fluorescent indication $UV_{254}$ (Macherey-Nagel) were used to monitor the progress of the reactions. The products were purified by column chromatography over silica gel (pore size 60 Å, 230-400 mesh particle size, 40-63 μm particle size, Sigma-Aldrich). $^1H$ and $^{13}C$ NMR spectra were recorded on a Bruker Avance 300 MHz (AV300) spectrometer using chloroform-d as a solvent. Electrospray ionization (ESI) high resolution mass spectra were recorded on a Finnigan LCQ mass spectrometer.

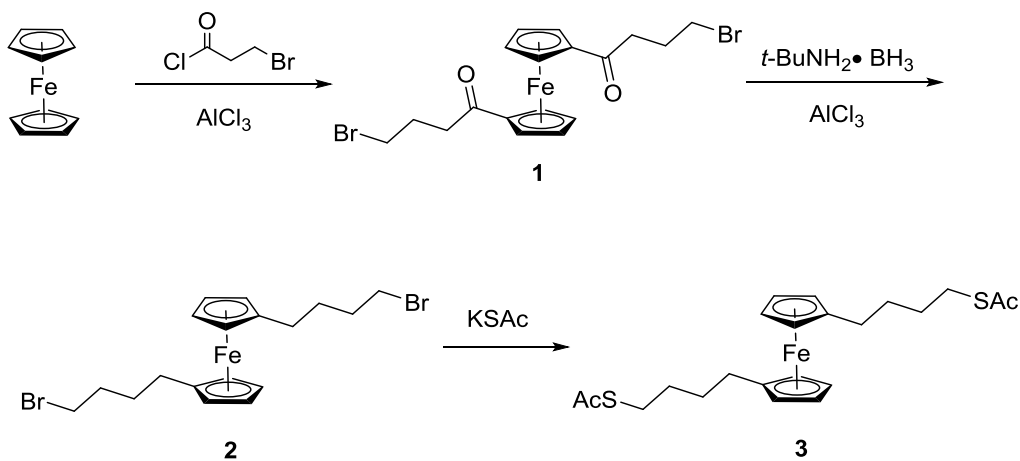

**Scheme 1.** Synthetic route to prepare AcS(CH$_2$)$_4$Fc(CH$_2$)$_4$SAc in three steps.

*Compound 1.* We followed a acylation procedure described in the literature [5] to synthesize ω-bromo aliphatic-1,1′-diacylferrocenes. In a 100 ml Schlenk flask, ferrocene (1.3 g, 7 mmol) and anhydrous AlCl$_3$ (2.8 g, 21 mmol) were dissolved in anhydrous DCM (30 mL), next a solution of 4-bromobutanoyl chloride (2 mL, 17 mmol) in anhydrous DCM (25 mL) was added to the reaction mixture dropwise at room temperature. The reaction mixture was stirred for 3 h at room temperature under a nitrogen atmosphere. After addition of deionized water (30 mL), the reaction mixture was stirred for an additional 10 min. The dark red colored organic layer was separated from the blue colored aqueous layer. The aqueous layer was extracted three times with DCM (25 mL) and the combined organic layers were washed with saturated sodium chloride, dried over sodium sulfate, filtered, and concentrated using rotary evaporation. The crude product was purified by column chromatography (hexane/DCM = 1:3) to yield the product (1,5 g, 45% yield) as a dark red oil.

$^1$H NMR (CDCl$_3$, 300 MHz) δ 4.83 (s, 4H), 4.54 (s, 4H), 3.58 (t, 4H, J=6.0 Hz), 2.88 (t, 4H, J=6.6 Hz), 2.26 (m, 4H); $^{13}$C NMR (CDCl$_3$, 75 MHz) δ 202.1, 80.1, 73.6, 70.6,

37.5, 33.8, 26.6; ESI HRMS *m/z* calcd for $C_{18}H_{21}FeBr_2O_2$ 482.9253, found 482.9249 $(M^++H)$.

*Compound 2.* The reduction was performed according to a literature [6] reported procedure. Borane-tert-butylamine complex (1.56 g, 18 mmol) in anhydrous DCM (50 mL) was added at 0 °C to a suspension of $AlCl_3$ (1.2 g, 9 mmol) in anhydrous DCM (50 mL), the resulting mixture was allowed to stir at 0 °C for 1h until a clear solution was obtained. A solution of compound **1** (1.45 g, 3 mmol) in anhydrous DCM (25 mL) was added dropwise. The reaction mixture was stirred at 0 °C for 2 h and then hydrolyzed with deionized water (30 mL). The aqueous layer was extracted three times with DCM (25 mL) and the combined organic layers were washed with 0.1 M HCl, water and saturated sodium chloride, dried over sodium sulfate, filtered, and concentrated using rotary evaporation. The crude product was purified by column chromatography (hexane/DCM = 10:1) to provide product (1.2 g, 88% yield) as a yellow oil.

$^1$H NMR (CDCl$_3$, 300 MHz) δ 4.07 (s, 8H), 3.41 (t, 4H, J=6.6 Hz), 2.29 (t, 4H, J=6.9 Hz), 1.88 (t, 4H, J=6.9 Hz), 1.65 (m, 4H); $^{13}$C NMR (CDCl$_3$, 75 MHz) δ 89.3, 69.3, 68.5, 33.8, 32.5, 29.6, 28.5; ESI HRMS *m/z* calcd for $C_{18}H_{24}FeBr_2$ 453.9589, found 453.9596 $(M^+)$.

*Compound 3.* In a 100 ml Schlenk flask, potassium thioacetate (0.72g, 6.25 mmol) and compound 2 (1.15 g, 2.5mmol) were dissolved in anhydrous THF (50 mL). The solution was refluxed for 2 hour, and then allowed to cool to room temperature overnight. The resulting mixture was poured into water and extracted with ethyl acetate. The combined organic layers were washed with saturated sodium chloride, dried over sodium sulfate, filtered, and concentrated using rotary evaporation. The crude product was purified by column chromatography (hexane/DCM = 3:1) to provide product (1 g, 95% yield) as a yellow oil. $^1$H NMR (CDCl$_3$, 300 MHz) δ 4.04 (s, 8H), 2.87 (t, 4H, J=6.9 Hz), 2.32 (s,

6H), 2.27 (t, 4H, J=7.2 Hz), 1.58 (m, 8H); $^{13}$C NMR (CDCl$_3$, 75 MHz) δ 195.9, 89.4, 69.2, 68.3, 30.6, 30.3, 29.3, 28.9, 28.8; ESI HRMS *m/z* calcd for C$_{22}$H$_{31}$FeO$_2$S$_2$ 447.1109, found 447.1107 (M$^+$+H).

## 6. References

[1] Tao, N. J. Electron transport in molecular junctions. *Nature Nanotech*. **1,** 173–181 (2006).

[2] O'Neill, K., Osorio, E. A. & van der Zant, H. S. J. Self-breaking in planar few-atom Au constrictions for nanometer-spaced electrodes. *Appl. Phys. Lett*. **90,** 133109 (2007).

[3] de Leon, N. P., Liang, W., Gu, Q. & Park, H. Vibrational excitation in single-molecule transistors: deviation from the simple Franck-Condon prediction. *Nano Lett*. **8,** 2963–2967 (2008).

[4] Hansen, W. *et al.* Zeeman bifurcation of quantum-dot spectra. *Phys. Rev. Lett*. **62,** 2168–2171 (1989).

[5] Vulugundam, G., Kumar, K., Kondaiah, P. & Bhattacharya, S. Efficacious redox-responsive gene delivery in serum by ferrocenylated monomeric and dimeric cationic cholesterols. *Org. Biomol. Chem.* **13,** 4310-4320 (2015).

[6] Gharib, B. & Hirsch, A. Synthesis and characterization of new ferrocene-containing ionic liquids. *Eur. J. Org. Chem.* **19,** 4123-4136 (2014).